\documentstyle[12pt,aasms4,flushrt]{article}

\lefthead{GALLI ET AL.}
\righthead{MAGNETIZED POLYTROPIC CLOUDS}

\begin{document}

\title{SCALE-FREE EQUILIBRIA \\
OF ISOPEDIC POLYTROPIC CLOUDS}

\author{D.~Galli$^1$, S.~Lizano$^2$, Z.~Y.~Li$^3$, F.~C.~Adams$^4$, and 
F.~H.~Shu$^5$}
\affil{$^1$~Osservatorio di Arcetri, Largo E. Fermi 5, 50125, Firenze, Italy}
\affil{$^2$~Instituto de Astronom\'\i a, UNAM, Apdo. 70264, 04510 M\'exico D. 
F., M\'exico}
\affil{$^3$~Astronomy Department, University of Virginia, Charlottesville, VA 
22903}
\affil{$^4$~Physics Depeartment, University of Michigan, Ann Arbor, MI 48109}
\affil{$^5$~Astronomy Department, University of California, Berkeley, CA 
94720-3411}

\begin{abstract}

We investigate the equilibrium properties of self-gravitating
magnetized clouds with polytropic equations of state with negative
index $n$.  In particular, we consider scale-free isopedic
configurations that have constant dimensionless spherical mass-to-flux
ratio $\lambda_r$ and that may constitute ``pivotal'' states for
subsequent dynamical collapse to form groups or clusters of stars.  For
given $\Gamma = 1 + 1/n$, equilibria  with smaller values of
$\lambda_r$ are more flattened, ranging from spherical configurations
with $\lambda_r = \infty$ to completely flattened states for $\lambda_r
= 1$.  For a given amount of support provided by the magnetic field as
measured by the dimensionless parameter $H_0$, equilibria with smaller
values of $\Gamma$ are more flattened. However, logatropic (defined by
$\Gamma =0$) disks do not exist. The only possible scale-free isopedic
equilibria with logatropic equation of state are spherical uniformly
magnetized clouds.

\end{abstract}


\section{INTRODUCTION}
\label{sec:Introd}

The stage leading up to dynamic collapse of a magnetically subcritical
cloud core to a protostar or a group of protostars is believed to be
largely quasi-static, if the responsible process is
ambipolar diffusion (e.g., Mestel \& Spitzer 1956,
\cite{Nakano79}, \cite{LS89},
\cite{Tomisaka89}, \cite{BM94}).\footnote{For example, as measured either
by the net accelerations or by the square of the inward flow speed
divided by the sound speed, the ambipolar-diffusion
models in Figures 3 and 6 of \cite{CM94} spend less than 0.1\%
of the total computed evolutionary time in states where even
a single grid point is more than 10\% out of mechanical
balance with self-gravity, magnetic forces, and thermal
pressure (see also Figs. 3 and 7 of Basu \& Mouschovias~1994 and 
Fig.~1 of Ciolek \& Koenigl~1998).}
To describe the transition between
quasi-static evolution by ambipolar diffusion and dynamical evolution
by gravitational collapse, \cite{LS96} introduced the idea of  a
pivotal state, with the scale-free, magnetostatic, density distribution
approaching $\rho \propto r^{-2}$ for an isothermal equation of state
(EOS) when the mass-to-flux ratio has a spatially constant value, a
condition that \cite{SL97} and \cite{LS97} termed ``isopedic''.
Numerical simulations of the contraction of magnetized clouds justify
the assumption of a nearly constant mass-to-flux ratio in the pivotal
core.\footnote{For example, inside the starred point where
\cite{CM94} consider the core to begin,
the mass-to-flux ratio varies in the last models of their Figures 3 and 6
by a factor of only 3 or 2 over a range
where the density varies by a factor $\sim 10^5$.
Outside the starred point, the mass-to-flux value
exhibits greater variation, but this occurs only because
\cite{CM94} impose starting
values for the mass-to-flux in the envelope that are
$\sim 2\times 10^{-2}$ times the critical value
(see also Figs. 4a and 8b in \cite{BM94}). Such small
ratios for the bulk of the mass of a molecular cloud
are probably ruled out by the Zeeman OH measurements
summarized by Crutcher (1998).}

The small dense cores of molecular clouds
that give rise to low-mass star formation
are effectively isothermal (\cite{MB83}; \cite{SAL87}).
The situation may be different for larger regions
that yield high-mass or clustered star formation.
It has often been suggested that the EOS relating the gas pressure $P$
to the mass density $\rho$ of interstellar clouds can be represented by
a polytropic relation $P \propto \rho^{1 + 1/n}$ with negative index
$n$.  \cite{Shu72} pointed out the utility of this idealization within the
context of the classic two-phase model of the diffuse interstellar
medium [\cite{Pikelner67}; \cite{FGH69}; \cite{SpitzerandScott69}],
while \cite{VialaandHoredt74} published extensive tables analyzing the
stability of non-magnetized, self-gravitating spheres of such gases.
\cite{Maloney88} examined the linewidth-size and density-size relations
of molecular clouds, first found by \cite{Larson81} and subsequently
studied by many authors [e.g., \cite{Leung82}, \cite{Torrelles83},
\cite{Dame86}, \cite{Falgarone92}, \cite{MieschB94}].  Maloney
pronounced the results consistent with the properties of negative index
polytropes.  For a polytropic EOS, the sound speed $c_s\equiv
(dP/d\rho)^{1/2}\propto \rho^{1/2n}$ increases with decreasing density
if $n<0$.  The latter behaviour may be compared with the empirical
linewidth-density relation for molecular clouds, $\Delta v\propto
\rho^{-q}$, with $q \simeq 0.5$ for low-mass cores (\cite{MF92}) and
$q \simeq 0.15$ for high-mass cores (\cite{CaM95}), implying that $n$
lies between $-1$ and $-3$, or that a static $\Gamma \equiv 1+1/n$ lies
between $0$ and 0.7.

The case $\Gamma=1/2$ is of particular relevance for the equilibrium
properties of molecular clouds. \cite{Wa44} found that the pressure of
Alfv\`en waves propagating in a stratified medium, $P_{\rm wave}\propto
|\delta{\bf B}|^2$, in the absence of damping obeys the simple
polytropic relation $P_{\rm wave} \propto \rho^{1/2}$, a consequence of
conservation of the wave energy flux $v_A |\delta{\bf B}|^2$. This
result was later derived more rigorously by \cite{W62} in the WKB
approximation for MHD waves propagating in mildly inhomogeneous media,
and, more recently by \cite{FA93} and \cite{McKZ95} in a specific
astrophysical context.  In numerical simulations of the same problem,
\cite{GO96} found indication of a much shallower relation ($\Gamma
\simeq 0.1$) for a self-gravitating medium supported by nonlinear
Alfv\`en waves.  On the other hand, for the adiabatic contraction of a
cloud supported by linear Alfv\`en waves, \cite{McKZ95} found a dynamic
$\gamma$ larger than 1.  \cite{Vazquez97} confirmed a similar behaviour
in numerical simulations of the gravitational collapse of clouds with
an initial field of hydrodynamic rather than hydromagnetic turbulence.

In the limit of $\Gamma\rightarrow 0$ (or $n\rightarrow -1$), the EOS
becomes ``logatropic,'' $P\propto \ln\rho$, a form first used by
\cite{LS89} to mimic the nonthermal support in molecular clouds
associated with the observed supersonic linewidths.  The sound speed
associated with the nonthermal contribution, $c_s = (d P / d
\rho)^{1/2} \propto \rho^{-1/2}$ becomes important at the low densities
characteristic of molecular cloud envelopes (as contrasted with the
cloud cores) since the thermal contribution is independent of density
if the temperature $T$ remains constant.  This nonthermal contribution
decreases with increasing density and will become subsonic at high
densities as recently observed in the central regions of
dense cores (\cite{BG98}).
\cite{McLP96} and \cite{McLP97} have
modeled the equilibrium and collapse properties of unmagnetized,
self-gravitating, spheres with a pure logatropic EOS and claim to find
good agreement with observations.

Adams, Lizano, \& Shu (in 1987) independently obtained the similarity solution
for the gravitational collapse of an unmagnetized singular logatropic
sphere (SLS), but they chose not to publish their findings until they
had learned how to magnetize the configuration in a nontrivial way (see
the reference to this work in Fuller \& Myers~(1992), who considered the
practical modifications to the protostellar mass-infall rate introduced
by ``nonthermal'' contributions to the support against self-gravity).
Magnetization constitutes an important program to carry out if we try
to justify a nonthermal EOS as the result of a superposition of
propagating MHD waves (see also Holliman \& McKee 1993).
In this paper, we extend the study of Li \& Shu
(1996) to include the isopedic magnetization of pivotally
self-gravitating clouds with a polytropic equation of state.  As a
by-product of this investigation, we obtain the unanticipated and
ironic result that the only way to magnetize a singular logatropic
configuration and maintain a scale-free equilibrium is to do it
trivially, i.e., by threading the SLS with
straight and uniform field lines (see \S 6).

A basic consequence of treating the turbulence as a scalar
pressure, coequal to the thermal pressure except for
satisfying a different EOS, is that we do not change the
basic topology of the magnetic field.  This assumption
may require reassessment if MHD turbulence enables
fast magnetic reconnection
(\cite{VL99}) and allows the magnetic fields
of highly flattened cloud cores (Mestel \& Strittmatter 1967,
\cite{BM96})
or pseudodisks (\cite{GS93b}) to disconnect from their
background.  Recent MHD simulations carried out
in multiple spatial dimensions
(e.g., \cite{sto98}; \cite{mac98}; \cite{ost99}; \cite{pad99})
find turbulence in strongly magnetized media to decay
almost as fast as in unmagnetized media.
Such decay may be responsible for accelerating
molecular cloud core formation above simple
ambipolar diffusion rates (\cite{Nakano98}, \cite{ML98}, \cite{shuetal99}).
Although this result also cautions against treating turbulence
on an equal footing as thermal pressure, we attempt
a simplified first analysis that includes
magnetization to assess the resulting configurational changes
when we adopt an alternative EOS for the pivotal state.
In particular, different power-law dependences of the radial
density profile translate immediately to different
time dependences in the mass-infall rate for the
subsequent inside-out collapse (\cite{Cheng78}, McLaughlin \& Pudritz~1997).

The paper is organized as follows.  In \S 2 we formulate the equations
of the scale-free problem and show that each solution depends only on
the polytropic exponent $\Gamma$ and a nondimensional parameter $H_0$
related to the cloud's morphology.  In \S 3 we present the numerical
results.  In \S 4, \S 5, and \S 6 we discuss the limiting form of the
solutions. Finally, in \S 7 we give our conclusions and discuss the
possible implications of our results for star formation and the
structure of giant molecular clouds.

\section{SELF-SIMILAR MAGNETOSTATIC EQUILIBRIUM EQUATIONS}

To begin, we generalize the singular polytropic sphere in the same way
that \cite{LS96} generalized the singular isothermal sphere (SIS).  In the
absence of an external boundary pressure, the only place the pressure $P$ enters 
in
the equations of magnetostatic equilibrium is through a gradient.
Consider then the polytropic relation
\begin{equation}
\label{dpdrho}
{dP\over d\rho}=K\rho^{-(1-\Gamma)}.
\end{equation}
By integrating equation (\ref{dpdrho}) we recover for $\Gamma = 1$ the
isothermal EOS, $P= K \rho$, (where $K$ is the square of the isothermal
sound speed) and for $\Gamma = 0$ the logatropic EOS, $P = K \ln\rho$.

We adopt axial symmetry in spherical coordinates and
consider a poloidal magnetic field given by
\begin{equation}
\label{bfield}
{\bf B} = {1 \over 2 \pi} \nabla \times \left( {\Phi \over r \sin \theta} \hat 
e_\phi \right),
\end{equation}
where $\Phi(r,\theta)$ is the magnetic flux.
Force balance along field lines requires
\begin{equation}
V+{1 \over \Gamma - 1} K\rho^{-(1 - \Gamma) }=h(\Phi),
\end{equation}
where $V$ is the gravitational potential and $h(\Phi)$ is the Bernoulli 
``constant'' along the
field line $\Phi =$ constant.
Poisson's equation now reads
\begin{equation}
\label{along}
{1\over r^2}{\partial\over\partial r}\left[r^2\left({dh\over d\Phi}
{\partial\Phi\over\partial r}- K\rho^{-(2-\Gamma)}{\partial\rho\over\partial r}
\right)\right]+{1\over r^2\sin^2\theta}
{\partial\over\partial\theta}\left[\sin\theta\left({dh\over d\Phi}
{\partial\Phi\over\partial\theta}-K\rho^{-(2-\Gamma)}
{\partial\rho\over\partial\theta}\right)\right]=4\pi G\rho;
\end{equation}
whereas force balance across field lines reads
\begin{equation}
\label{across}
{1\over 16\pi^3r^2\sin^2\theta}\left({\partial^2\Phi\over\partial r^2}+
{1\over r^2}{\partial^2\Phi\over\partial\theta^2}-{\cot\theta\over r^2}
{\partial\Phi\over\partial\theta}\right)=-\rho{dh\over d\Phi}.
\end{equation}

We look for scale-free solutions of the above equations by
nondimensionalizing and separating variables:
\begin{mathletters}
\begin{equation}
\rho=\left({K\over 2\pi Gr^2}\right)^{1/(2-\Gamma)}R(\theta),
\end{equation}
\begin{equation}
\Phi=4\left({\pi^{3-2\Gamma}Kr^{4-3\Gamma}\over G^{\Gamma/2}}
\right)^{1/(2-\Gamma)}\phi(\theta),
\end{equation}
\begin{equation}
\label{nondim}
{dh\over d\Phi}=H_0\left(2^{3\Gamma-2}KG^{2-2\Gamma}\over \pi^{1-\Gamma}
\Phi^{2-\Gamma}\right)^{1/(4-3\Gamma)},
\end{equation}
\end{mathletters}
where $H_0$ is a dimensionless constant that measures the deviation
from a force free magnetic field, and $R(\theta)$ and $\phi(\theta)$
are dimensionless functions of the polar angle $\theta$.\footnote{These
definitions are not applicable for $\Gamma = 4/3$ or $\Gamma=2$.}
These assumptions imply that the equilibria will have spatially
constant mass-to-flux ratios (see below).  Substitution of equation
(\ref{nondim}) into equations (\ref{along}) and (\ref{across}) yields
$$
{1\over\sin\theta}{d\over d\theta}\left\{\sin\theta
\left[A_\Gamma H_0\phi^{-(2-\Gamma)/(4- 3\Gamma)}\phi^\prime-
R^{-(2-\Gamma)}R^\prime\right]\right\}=
$$
\begin{equation}
\label{alongn}
2\left[R-{(4-3\Gamma) \over(2-\Gamma)^2}
R^{-(1-\Gamma)}-\left({4 - 3\Gamma\over 2 - \Gamma}\right)^2B_\Gamma H_0
\phi^{2(1-\Gamma)/(4-3\Gamma)}\right],
\end{equation}
and
\begin{equation}
\label{acrossn}
{1\over\sin^2\theta}\left[\phi^{\prime\prime}-\cot\theta\phi^\prime+
{2(4-3\Gamma)(1-\Gamma)\over(2-\Gamma)^2}\phi\right]=-C_\Gamma H_0R
\phi^{-(2-\Gamma)/(4-3\Gamma)},
\end{equation}
where a prime denotes differentiation with respect to $\theta$, and
\begin{mathletters}
\begin{equation}
A_\Gamma=2^{\Gamma(3-2\Gamma)/(4-3\Gamma)(2-\Gamma)},
\end{equation}
\begin{equation}
B_\Gamma=2^{-(1-\Gamma)(8-5\Gamma)/(4-3\Gamma)(2-\Gamma)},
\end{equation}
\begin{equation}
C_\Gamma=2^{-\Gamma(1-\Gamma)/(4-3\Gamma)(2-\Gamma)}.
\end{equation}
\end{mathletters}

In particular, for $H_0=0$, eq.~(\ref{alongn}) gives the dimensionless density
for the non-magnetized singular polytropic sphere 
\begin{equation}
\label{nonmagR}
R = \left[ {4-3\Gamma\over (2-\Gamma)^2}\right]^{1/(2-\Gamma)},
\end{equation}
whereas eq.~(\ref{acrossn}) implies $\Phi=0$ for $0 < \Gamma \le 1$, in
order to satisfy the boundary conditions eq.~(\ref{bc}).  In this case,
the mass-to-flux ratio $\lambda_r$ is infinite. However, for $\Gamma=0$, 
eq.~(\ref{acrossn})
admits also the analytic solution of $\Phi \propto r^2 \sin^2\theta$
corresponding to a straight and uniform field, while the density
function is $R(\theta) = 1$.  Therefore, a spherical logatropic scale
free cloud can be magnetized with a uniform magnetic field of any
strength, and any value of the spherical mass to flux ratio is
allowed.\footnote{In this case, $\lambda_r^2=2\mu^2=[2\phi(\pi/2)^2]^{-1}$.}

For arbitrary values of $\Gamma$ and $H_0$ the ordinary differential 
equations (ODEs) (\ref{alongn}) and
(\ref{acrossn}) are to be integrated subject to the two-point boundary 
conditions (BCs):  
$$
\lim_{\theta\rightarrow 0}\sin\theta\left[A_\Gamma
H_0\phi^{-(2-\Gamma)/(4-3\Gamma)} \phi^\prime-R^{-(2-\Gamma)}
R^\prime\right]=0, 
$$ 
\begin{equation} 
\label{bc} 
\phi(0)=0,~~~\phi^\prime(\pi/2)=0,~~~R^\prime(\pi/2)=0.  
\end{equation} 
The first BC implies that there is no contribution from the polar axis
to the mass inside a radius $r$.  The second BC comes from the
definition of magnetic flux, i.e. no trapped flux at the polar axis.
The last two BCs imply no kinks at the midplane.

The equilibria are characterized by:

\noindent ({\em a}\/) the spherical mass-to-flux ratio, \footnote{The
standard mass-to-flux ratio $\lambda = 2 \pi G^{1/2} M(\Phi)/ \Phi$ is
not defined for the polytropic scale free magnetized equilibria because
the integral $\int_0^{\pi/2} R(\theta) \phi(\theta)^{-1}\sin\theta d\theta$ 
diverges since it can be shown that $R(\theta=0) \ne 0$ for $\Gamma <1$.}
\begin{equation} 
\label{lambdar} 
\lambda_r\equiv 2 \pi
G^{1/2} { M(r) \over \Phi(r,\pi/2)} = 2^{(1-\Gamma)/(2-\Gamma)} \left(
{2 -\Gamma \over 4 - 3\Gamma} \right ) {1 \over \phi(\pi/2)}
\int_0^{\pi/2}R(\theta)\sin\theta d\theta , 
\end{equation} 
where $M(r)$ is the mass enclosed within a radius $r$;

\noindent ({\em b}\/) the factor $D$ by which the average density is enhanced
over the non-magnetized value because of the extra support provided
by magnetic fields,
\begin{equation}
\label{D}
D \equiv \left[ {4-3\Gamma\over (2-\Gamma)^2}\right]^{-1/(2-\Gamma)}
\int_0^{\pi/2}R(\theta)\sin \theta \, d\theta ,
\end{equation}
which is equal to $1$ if $H_0=0$ (see eq.~[\ref{nonmagR}]);

\noindent ({\em c}\/) the sound speed,
\begin{equation}
\label{cs}
c_s^2 = \left( 2 \pi G r^2 \right)^{(1-\Gamma)/(2-\Gamma)} K^{1/(2-\Gamma)} 
R(\theta)^{-(1-\Gamma)};
\end{equation}
and ({\em d}\/) the Alfv\`en speed 
\begin{equation}
v_A^2=2^{\Gamma/(2-\Gamma)} \left( 2 \pi G r^2 \right)^{(1-\Gamma)/(2-\Gamma)} 
K^{1/(2-\Gamma)}
\left[ (\phi^\prime)^2 + \left({4-3\Gamma \over 2-\Gamma}\right)^2 \phi^2 
\right]
{1 \over R(\theta) \sin^2\theta}. 
\end{equation}
Both the sound speed and the Alfv\`en speed scale as $r^0$ for $\Gamma
= 1$, and $r^{1/2}$ for $\Gamma = 0$; for other values of $\Gamma$, the
exponent of $r$ lies between these two values.

It is also of interest to define the ratio $\mu^2$ of the square of the sound
speed and the square of the Alfv\`en speed, each weighted by the
density, which is a physical quantity that can be compared with
observations:
\begin{equation}
\mu^2 = {\int_0^{\pi/2} 
c_s^2\rho\sin\theta d\theta 
\over \int_0^{\pi/2} v_A^2  \rho\sin\theta d\theta} = 
2^{-\Gamma/(2-\Gamma)}{\int_0^{\pi/2} R(\theta)^{\Gamma} \sin\theta d\theta
\over \int_0^{\pi/2} \left[ (\phi^\prime)^2 + 
\left({4-3\Gamma \over 2-\Gamma}\right)^2 \phi^2 \right]
/\sin\theta d \theta}.
\end{equation}

If $c_s$ represents only the thermal sound speed, then the
observational summary given by Fuller \& Myers~(1992) would imply that
$\mu^2 \sim 1$ in the quiet low-mass cores of GMCs, whereas $\mu^2 \sim
10^{-2}$ in their envelopes.  If we include in $c_s$, however, the
turbulent contribution, then the turbulent speed is likely to be
sub-Alfv\'enic or marginally Alfv\'enic, and $\mu^2 \lesssim 1$
everywhere is probably a better characterization of realistic clouds.

\section{RESULTS}
\label{sec:results}

To obtain an equilibrium configuration for given values of $\Gamma$ and
$H_0$, equations (\ref{alongn}) and (\ref{acrossn}) are integrated
numerically.  The integration is started at $\theta=0$ using the
expansions:  $\phi = a_0 \xi^2+ \ldots$, $ R = b_0 + b_2
\xi^{4(1-\Gamma)/ (4-3\Gamma)}+ \ldots$, with $\xi=\sin\theta$, and
$b_2 = [(4-3\Gamma)/2(1-\Gamma)] A_\Gamma H_0
a_0^{2(1-\Gamma)/(4-3\Gamma)} b_0^{2-\Gamma}$.  The values of $a_0$ and
$b_0$ are varied until the two BCs at $\theta = \pi/2$ (eq.~\ref{bc}),
are satisfied.  For flattened equilibria (see below) it is more
convenient to start from $\theta=\pi/2$, where the BCs $\phi^\prime
(\pi/2)=0$ and $R^\prime(\pi/2)=0$ are imposed, and integrate toward
$\theta=0$.  The values of $\phi(\pi/2)$ and $R(\pi/2)$ are then varied
until a solution is found that satisfies the two BCs at $\theta = 0$.

Figure~1 shows the resulting flux and density functions $\phi(\theta)$
and $R(\theta)$ computed for $H_0 = 0.5$ and values of $\Gamma$ between
0.2 and 1.  We reproduce the results of \cite{LS96} for $\Gamma = 1$,
which is the only case that obtains perfect toroids (i.e., $R[\theta=0]
=0$); models with $\Gamma < 1$ have nonzero density at the polar
axis.  Figure~2 shows the corresponding density contours and magnetic
field lines.  In the limit $\Gamma\rightarrow 0$, independent of $H_0$
as long as it is nonzero, the pivotal configuration becomes thin disks
with an ever increasing magnetic field strength.  Table 1 shows the
spherical mass to flux ratio $\lambda_r$, the overdensity parameter
$D$, and the ratio of the square of the sound and Alfv\`en speeds
$\mu^2$.  This table shows that, for fixed $H_0$, $\mu^2$ decreases as
$\Gamma$ decreases because the magnetic field becomes stronger. For the
same reason $D$ increases.  In contrast, $\lambda_r$ goes through a
minimum as $\Gamma$ decreases.  Figures~1 and 2 demonstrate that for
$\Gamma \rightarrow 0$ (the logatropic limit), $H_0$ is not a measure
of the strength of the magnetic fields since $\phi$ diverges as
$R(\theta) \rightarrow \delta(\pi/2-\theta)$ (see \S 5 below).

For fixed $\Gamma$, a sequence from small $H_0$ to large $H_0$
progresses through configurations of increasing support by magnetic
fields, as demonstrated explicitly for the isothermal case by
\cite{LS96}.  This behavior is illustrated here for the $\Gamma = 1/2$
case in Figure~3, which shows the density contours and magnetic field
lines corresponding to values of $H_0$ from 0.05 to 1.5.  Table 2 shows
the corresponding values of $\lambda_r$, $D$, and $\mu^2$.  For small
$H_0$, the equilibria have nearly spherically symmetric isodensity
contours and weak quasiuniform magnetic fields that provide little
support against gravity.  With increasing $H_0$, the pivotal
configurations flatten.  The case $H_0=1.5$ is already quite disklike:
the pole to equator density contrast is $R(\pi/2)/R(0) \simeq 10^6$.
For a thin disk, the analysis of \cite{SL97} demonstrates that magnetic
tension provides virtually the sole means of horizontal support against
self-gravity, with gas and magnetic pressures being important only for
the vertical structure.  In the limit of a completely flattened disk
($H_0\rightarrow \infty$), $\lambda_r \rightarrow 1$ independent of the
detailed nature of the gas EOS (see next section).  Table 2 shows the
spherical mass to flux ratio $\lambda_r$, the overdensity parameter
$D$, and the ratio of the square of the sound and Alfv\`en speeds
$\mu^2$. Again $D$ increases monotonically and $\mu^2$ decreases
monotonically as the magnetic support increases with $H_0$, while
$\lambda_r$ goes through a minimum and tends to 1 for large $H_0$.

Since the mass-to-flux ratio $\lambda_r$ is a fundamental quantity that
will not change unless magnetic field is lost by ambipolar diffusion,
in Figure~4 we consider sequences where $\lambda_r$ is held fixed, but
$\Gamma$ is varied.  This Figure shows the locus of the set of
equilibria with $\lambda_r =0.95,1,$ and $2$ in the ($H_0$, $\Gamma$)
plane.  Equilibria with $\lambda_r < 1$ are highly flattened when
$\Gamma \rightarrow 0$ even for small but fixed values of $H_0$ (see \S
6).  In fact, to obtain incompletely flattened clouds when one takes
the limit $\Gamma \rightarrow 0$, one also needs simultaneously to
consider the limit $H_0\rightarrow 0$.  Unfortunately, because both
the density and the strength of the magnetic field at the midplane
diverge as the equilibria become highly flattened, we are unable to
follow numerically the limit $\Gamma \rightarrow 0$ to verify if these
sequences of constant $\lambda_r < 1$ will hook to a finite value in
the $H_0$ axis, or will loop to $H_0 =0$, consistent with our
demonstration in \S4 that flattened disks do not exist in the
logatropic limit.\footnote{As the equilibria flatten due to
either small $\Gamma$ or large $H_0$, it becomes necessary to determine
the constants of the expansions of $R(\theta)$ and $\phi(\theta)$ near
the origin with prohibitively increasing accuracy.}

%Indeed, the fact
%that the last computed model in Table 1 with $\Gamma = 0.2$ has
%$\lambda_r = 0.992$, seems to indicate that the sequences of constant
%$\lambda_r$ loop back to $H_0 = 0$.

We speculate that the results for $\lambda_r < 1$ have the following
physical interpretation.  According to the theorem of Shu \& Li (1997),
only if $\lambda$ itself rather than $\lambda_r$ is less than unity,
the magnetic field is strong enough overall to prevent the
gravitational collapse of a highly flattened cloud.  However, for
moderate $H_0$ and $\Gamma$ when $\lambda_r < 1$, even the singular
equilibria are probably magnetically subcritical, since there can be
little practical difference between the spherical mass-to-flux ratio
$\lambda_r$ and the ``true'' mass-to-flux ratio $\lambda$ for highly
flattened configurations.  The latter is formally infinite when $\Gamma
< 1$ only because the mass column goes to zero a little slower than the
field column when we perform an integration along the central field
line (see footnote 3).  In this interpretation, subcritical scale-free
clouds with $\lambda_r < 1$ and intermediate values of $\Gamma$ can
become highly flattened because magnetic tension supports them
laterally against their self-gravity while the soft EOS does not
provide much resistance in the direction along the field lines.  The
squeezing of the cloud toward the midplane is compounded by the {\it
confining} pressure of bent magnetic field lines that exert pinch
forces in the vertical direction.  Both the magnetic tension and the
vertical pinch of magnetic pressure disappear when the field lines
unbend, as they must to maintain the scale-free equilibria in the limit
$\Gamma \rightarrow 0$ (see below).  As a consequence, logatropic
configurations become spherical for any value of $\lambda_r$.  We leave
as an interesting problem for future elucidation the determination
whether there is still a threshold in $\lambda_r$ below which the SLS,
embedded with straight and uniform field lines, will not collapse
dynamically.

\section{THE THIN DISK LIMIT ($H_0 \gg 1$)}

In the limit $H_0\gg 1$, the cloud flattens to a thin disk for any
$\Gamma\le 1$.  Dominant balance arguments applied to the two ODEs of
the problem reveal the following asymptotic behaviour:\footnote{These
expansions are not valid for $\Gamma = 1$. See Li \& Shu~(1996) for the
correct asymptotic expansion in this case.}
\begin{equation}
\label{RH0}
R(\theta) \rightarrow R_0 \delta(\theta -\pi/2) H_0^{(4-3\Gamma)/(2-\Gamma)}
+ s(\theta) H_0^{-(4-3\Gamma)/(2-\Gamma)(1-\Gamma)},
\end{equation}
\begin{equation}
\label{fH0}
\phi \rightarrow f(\theta) H_0^{(4-3\Gamma)/(2-\Gamma)}.
\end{equation}
To the lowest order in $H_0$ the equation of force balance along field 
lines (eq.~\ref{alongn}) becomes:
\begin{eqnarray}
\label{asymp}
& & {1\over\sin\theta}{d\over d\theta}\left[\sin\theta\left(A_\Gamma
f^{-(2-\Gamma)/(4-3\Gamma)}
f^{\prime}-s^{-(2-\Gamma)}s^{\prime}\right)\right]= \nonumber \\
& & -2\left[{4-3\Gamma\over (2-\Gamma)^2}s^{-(1-\Gamma)}
+\left({4-3\Gamma\over 2-\Gamma}\right)B_\Gamma 
f^{2(1-\Gamma)/(4-3\Gamma)}\right],
\end{eqnarray}
valid over the interval  $0\leq \theta < \pi/2$, plus the 
the integral constraint
\begin{equation}
\label{int1}
R_0 -{4-3\Gamma \over (2 - \Gamma)^2} \int_0^\pi s^{-(1-\Gamma)}\sin\theta
d\theta  
- \left({4 - 3 \Gamma \over 2 -\Gamma}\right)^2
B_\Gamma \int_0^\pi f^{2(1-\Gamma)/(4-3\Gamma)} \sin\theta d\theta=0,
\end{equation}
obtained by integrating eq.~(\ref{alongn}) from $\theta=0 $ to $\pi$, and
applying the first BC (eq.~\ref{bc}) on the polar axis.  

The constant $R_0$ is proportional to the surface density of the polytropic 
disks, given by 
\begin{equation}
\label{surfden}
\Sigma(r) \equiv \lim_{\epsilon\rightarrow 0} 
\int_{\pi/2-\epsilon}^{\pi/2+\epsilon}
\rho r \sin\theta d\theta \rightarrow \left({K \over 2 \pi G}\right)^{1 /(2 - 
\Gamma)} 
r^{-\Gamma/(2-\Gamma)} R_0 H_0^{(4-3\Gamma)/(2-\Gamma)},
\end{equation}
which, for $\Gamma = 1$ gives $\Sigma \rightarrow H_0 a^2/\pi G r$, as
found by \cite{LS96}.

Eq.~(\ref{acrossn}) expressing force balance across field lines reduces to 
\begin{equation}
{1\over\sin^2\theta}\left[f^{\prime\prime}-\cot\theta f^\prime+
\ell(\ell +1) f\right]=-C_\Gamma
f^{-(2-\Gamma)/(4-3\Gamma)}R_0\delta(\theta-\pi/2), 
\end{equation}
where the parameter $\ell$ is defined by 
\begin{equation} 
\ell(\ell +1)\equiv {2(4-3\Gamma) (1-\Gamma)\over (2-\Gamma)^2}.  
\end{equation}
This is equivalent to the equation for force free magnetic fields
\begin{equation} 
\label{forcefree} 
f^{\prime\prime} - \cot\theta
f^\prime + \ell(\ell +1) f = 0, 
\end{equation} 
valid over the interval
$0\leq \theta < \pi/2$, plus the condition 
\begin{equation}
\label{int2} 2 f^\prime(\pi/2) = C_\Gamma R_0
f(\pi/2)^{-(2-\Gamma)/(4-3\Gamma)}, 
\end{equation} 
obtained integrating eq.~(\ref{acrossn}) from $\pi/2 - \epsilon$ to
$\pi/2 + \epsilon$, and taking the limit $\epsilon\rightarrow 0$.  For
integer $\ell$, solutions of eq.~(\ref{forcefree}) regular at
$\theta=0$ are Gegenbauer polynomials of order $\ell$ and index
$\frac{1}{2}$, $C^{(\frac{1}{2})}_\ell$ (see e.g.  Abramowitz \&
Stegun~1965). In general, it can be shown (Chandrasekhar~1955) that any
axisymmetric force free field, separable in spherical coordinates, can
be expressed in terms of fundamental solutions whose radial dependence
is given by a combination of Bessel functions of fractional order, and
the angular dependence by Gegenbauer polynomials of index
$\frac{1}{2}$. In our case, the choice of $\Gamma$ determines a
particular exponent of the power-law for the radial part of the flux
function, and hence the corresponding value of $\ell$ (non-integer,
except for $\Gamma=0$ and 1).

Therefore, the magnetic field is force free everywhere except at the
midplane where $\rho \neq 0$ and the condition of force balance across
field lines has to be satisfied.  In the thin disk limit discussed
here, the boundary condition $\phi^\prime(\pi/2) =0$ is clearly not
fullfilled:  the kink of $\phi$ at the midplane provides the magnetic
support against self-gravity on the midplane. Currents must exist in
the disk to support these kinks.

With the definitions
$$
y(\theta) \equiv -A_\Gamma {4-3\Gamma\over 2} 
f(\theta)^{2(1-\Gamma)/(4-3\Gamma)}, ~~~~~ z(\theta)\equiv 
s(\theta)^{-(1-\Gamma)},
$$
eq.~(\ref{asymp})
transforms into
\begin{equation}
z^{\prime\prime}+\cot\theta z^{\prime}+\ell(\ell+1)z=
y^{\prime\prime}+\cot\theta y^{\prime}+\ell(\ell+1)y,
\end{equation}
which has the solution
$$
z(\theta)=y(\theta)+q(\theta),
$$
where $q(\theta)$ is a solution of the homogeneous equation 
\begin{equation}
\label{eqq}
q^{\prime\prime}+\cot\theta q^{\prime}+\ell(\ell+1)q=0.
\end{equation}
Therefore, 
\begin{equation}
\label{sol}
s(\theta)=\left[q-A_\Gamma {4-3\Gamma\over 2}f^{2(1-\Gamma)/(4-3\Gamma)}
\right]^{-1/(1-\Gamma)},
\end{equation}
and the integral constraint
eq.~(\ref{int1})  becomes
\begin{equation}
\label{qconst}
\int_0^{\pi/2} q(\theta)\sin\theta d\theta={(2-\Gamma)^2\over 2(4-3\Gamma)}R_0.
\end{equation}

The problem is thus reduced to the solution of the two homogeneous equations
eq.~(\ref{forcefree}) and eq.~(\ref{eqq}) for the functions $f(\theta)$
and $q(\theta)$ which are determined up to an arbitrary constant.
However, the two integral constraints that would have determined
these latter constants (eqs.~\ref{int2}, \ref{qconst}), contain the
additional unknown parameter $R_0$.  The system of equations is closed
by the requirement that
$$ \lim_{H_0 \rightarrow \infty} \lambda_r = 1.$$
Substituting eq.~(\ref{RH0}) and eq.~(\ref{fH0})
in eq.~(\ref{lambdar}), one obtains
$$  \lim_{H_0 \rightarrow \infty} \lambda_r =
2^{(1-\Gamma)/(2-\Gamma)} \left({2-\Gamma \over 4-3\Gamma} \right)
{R_0 \over 2f({\pi/ 2})} = 1,$$
i.e.,
\begin{equation}
\label{int3}
R_0 = 2^{1/(2-\Gamma)} \left( {4-3\Gamma \over 2-\Gamma} \right) f({\pi/ 2}),
\end{equation}
which gives the remaining condition.

Eq.~(\ref{forcefree}) and (\ref{eqq}) can be solved numerically by starting
the integration at $\theta=0$ with the 
series expansions: 
$q(\theta)=q_0 [ 1-{1 \over 4}\ell(\ell+1)\theta^2 + \ldots ]$, 
and $f(\theta) = f_0\left\{ \theta^2 -{1 \over 8} [\ell(\ell+1)+{2\over 3}] 
\theta^4 + \ldots\right\}$,
where $q_0$ and $f_0$ are arbitrary constants.\footnote{The two
original BCs on the function $R(\theta)$ are of little use here: the
one at $\theta=0$ reduces to the condition $\lim_{\theta\rightarrow 0}
(1-\Gamma)^{-1}\sin\theta q^\prime=0$, trivially satisfied; the second
BC, $R^\prime(\pi/2)=0$ cannot be applied because of the
$\delta$-function at $\pi/2$.}  The constants $q_0, f_0$ and $R_0$ are
then determined by the constraints expressed by
eqs.~(\ref{int1}),~(\ref{int2}),and~(\ref{int3}).

Figure~5 shows the functions $f(\theta)$ and $s(\theta)$ obtained for
$\Gamma=1/2$ and increasing values of $H_0$ from 0.4 to 1.5 compared
with the asymptotic expressions computed here. Already for $H_0 = 1.5$,
the actual $f(\theta)$ and $s(\theta)$ are very close to the 
corresponding asymptotic functions eq.~(\ref{RH0}) and eq.~(\ref{fH0}).
%Nevertheless it is important to
%point out that the asymptotic solution eq.~(\ref{sol}) is arrived only
%in the limit $H_0 \rightarrow \infty$, in such a way that $\lambda_r$
%is always finite, since the  the integral $\int_0^{\pi/2}
%H_0^{-1/(1-\Gamma)} r(\theta) \sin\theta d\theta$ must be finite.
Table 3 shows the value of the angle $\alpha$ of the magnetic field
with the plane of the disk, the flux function $f$ evaluated at
$\theta=\pi/2$ (indicative of the magnetic field stength), and the
surface density parameter $R_0$, as functions of $\Gamma$. The angle
$\alpha$ ranges from $45^\circ$ for the isothermal case $\Gamma = 1$ to
$90^\circ$ in the logatropic case $\Gamma =0$.  Correspondingly, the
magnetic flux in the disk and the surface density both diverge as
$\Gamma \rightarrow 0$ for any large but finite value of $H_0$.

\section{\bf THE QUASI-SPHERICAL LIMIT ($H_0 \ll 1$)}

For the isothermal case $\Gamma=1$, \cite{LS96} have shown how the
SIS is recovered for $H_0\ll 1$ from a family of
toroids with zero density on the polar axis.  For $\Gamma\ne 1$,
in the limit $H_0 \ll 1$, the asymptotic expansions are given by:
$$R(\theta) \rightarrow \left[ {4-3\Gamma \over (2 -\Gamma)^2} 
\right]^{1/(2-\Gamma)} 
+ p(\theta) H_0^{(4-3\Gamma)/(3-2\Gamma)} + \ldots$$
$$\phi = g(\theta) H_0^{(4-3\Gamma)/2(3-2\Gamma)} + \ldots.$$
To the lowest order in $H_0$, eqs.~(\ref{alongn}) and (\ref{acrossn}) become:
$$
{1\over\sin\theta}{d\over d\theta}\left\{\sin\theta
\left[A_\Gamma g^{-(2-\Gamma)/(4- 3\Gamma)} g^\prime-
{(2 -\Gamma)^2 \over 4 - 3 \Gamma} p^\prime\right]\right\}=
$$
\begin{equation}
2\left[(2-\Gamma)p-\left({4-3\Gamma \over 2-\Gamma}\right)^2 
B_\Gamma g^{2(1-\Gamma)/(4-3\Gamma)}\right],
\end{equation}
and
\begin{equation}
\label{fluxsmallh}
{1\over\sin^2\theta}\left[g^{\prime\prime}-\cot\theta g^\prime+
\ell(\ell+1) g\right]=-C_\Gamma
\left[{4-3\Gamma\over (2-\Gamma)^2}\right]^{1/(2-\Gamma)}
g^{-(2-\Gamma)/(4-3\Gamma)},
\end{equation}
The BC for the functions $p(\theta)$ and $g(\theta)$ are the same as
those for $R(\theta)$ and $\phi(\theta)$ in eq.~(\ref{bc}).

Figure~6 shows the convergence of the solutions of the full set of
equations (\ref{alongn}) and (\ref{acrossn}) obtained for $\Gamma=1/2$
and decreasing values of $H_0$ from 0.4 to 0.05, to the asymptotic
solutions obtained by integrating the equations above. Notice that $p(0) < 0$ 
and
$p(\pi/2) > 0$ showing that the sequence of equilibria with
$\Gamma=1/2$ originates from the corresponding unmagnetized spherical
state (eq.~\ref{nonmagR}) by reducing the density on the pole and
enhancing it on the equator. The same behaviour is found for any value
of $\Gamma$ in the range $0<\Gamma < 1$. For $\Gamma=1$, the function
$p(\theta)$ diverges at $\theta=0$, indicating that this expansion is
not appropriate in the isothermal case, as in the case $H_0\gg 1$. For
the same reason, the expansion also fails for $\Gamma=0$, since both
$g(\theta)$ and $p(\theta)$ diverge on the equatorial plane.

These flattened configurations are supported by magnetic and gas
pressure against self-gravity. The intensity of the magnetic field can
become very high even though $H_0$ is small, because the latter
parameter measures not the field strength but the deviations from a
force free field (see eq.~\ref{nondim}).

\section{\bf THE LOGATROPIC LIMIT ($\Gamma \rightarrow 0$).}

We consider in this section the logatropic limit $\Gamma\rightarrow
0$.  As anticipated in \S~2, for $\Gamma=0$ eq.~(\ref{alongn}) and
(\ref{acrossn}) admit the analytical solution $R=1$ and $\Phi\propto
r^2\sin^2\theta$ corresponding to a SLS threaded by a straight and
uniform magnetic field. This solution represents the only possible
scale-free isopedic configuration of equilibrium for a magnetized cloud
with a logatropic EOS. To show this, we use the results of \S~4 and
\S~5 for $H_0\gg 1$ and $H_0\ll 1$ to find the limit of the equilibrium
configurations for $\Gamma\rightarrow 0$ and fixed (small or large)
values of $H_0$.

In the limit $H_0 \gg 1$, $\Gamma \rightarrow 0$, analytic solutions to
equations ~(\ref{forcefree}) and ~(\ref{eqq}) exist.  The magnetic
field tends to become uniform and straight, $f(\theta) \rightarrow
f(\pi/2)\sin^2\theta$, but $f(\pi/2)$ diverges, as shown in Table~3,
and therefore $s(\theta=\pi/2)$ also diverges (see eq.~\ref{sol}).
Eq.~(\ref{surfden}) shows in this limit that the surface density
$\Sigma$ is independent of $r$, therefore, no pressure gradients can be
exerted in the horizontal direction. The value $\Sigma = (K/2 \pi
G)^{1/2} R_0 H_0^2 $  diverges as $\Gamma \rightarrow 0$ for any value
of $H_0$ because $\lim_{\Gamma \rightarrow 0} R_0 = \infty$ (see
Table~3).  The magnetic flux threading the disk, $\phi = 2^{-3/2}  R_0
H_0^2 r^2 \sin^2\theta$, becomes infinite in order to keep the mass to
flux ratio $\lambda_r $ equal to 1.  Therefore, the limiting
configuration approaches a uniform disk with infinite surface density,
threaded by an infinitely strong uniform and straight magnetic field.

If we now examine the case $H_0 \ll 1$, in the limit $\Gamma
\rightarrow 0$, it is easy to show from eq.~(\ref{fluxsmallh}) that the
magnetic field tends to become uniform, $g(\theta) \rightarrow
g(\pi/2)\sin^2\theta$, but $\lim_{\Gamma\rightarrow
0}g(\pi/2)=\infty$.  Consequently, the density function
$p(\theta)$ also diverges in $\theta= \pi/2$, and the configuration again
approaches a thin disk threaded by an uniform, infinitely strong,
magnetic field.

We conclude that scale-free logatropic clouds cannot exist as
magnetostatic disks except in some limiting configuration.  In the
absence of such limits, the equilibria are spherical and can be
magnetized only by straight and uniform field lines; i.e., the magnetic
field is force-free and therefore given by $H_0=0$.  The inside-out
gravitational collapse of such a SLS would still proceed self-similarly
as in the solution of McLaughlin \& Pudritz (1997), but the frozen-in
magnetic fields would yield a dependence with polar angle that
eventually produces a pseudodisk (Galli \& Shu 1993a,b; Allen \& Shu
1998a).

\section{SUMMARY AND DISCUSSION}

We have solved the scale-free equations of magnetostatic equilibrium of
isopedic self-gravitating polytropic clouds to find pivotal states that
represent the initial state for the onset of dynamical collapse, as
first proposed by \cite{LS96} for isothermal clouds.  Compared to
unmagnetized equilibria, the magnetized configurations are flattened
because of magnetic support across field lines.  The degree of this
support is best represented by the ratio of the square of the sound to
Alfv\`en speeds $\mu^2$, or the overdensity parameter $D$, since they
are always monotonic functions of $H_0$ and $\Gamma$.

Configurations with $\Gamma = 1$ become highly flattened as the
parameter $H_0$ increases.  When $\Gamma < 1$ (softer EOS) the
equilibria get flattened even faster at the same values of $H_0$, since
along field lines there is less support from a soft EOS than for a
stiff one.  However, it seems that in the logatropic limit flattened
disks do not exist: the singular scale-free equilibria can only be
spherical uniformly magnetized clouds.  Figure~7 shows a schematic
picture of the $(H_0, \Gamma)$ plane indicating the topology of the
solutions for scale free magnetized isopedic singular self-gravitating
clouds.

In self-gravitating clouds, the joint compression of matter and field
is often expressed as producing an expected relationship:  $B\propto
\rho^\kappa$, with different theorists expressing different preferences
for the value of $\kappa$ (e.g., Mestel 1965, Fiedler \& Mouschovias
1993).  No local (i.e., point by
point) relationship of the form $B\propto \rho^\kappa$ holds for the
scale-free equilibria studied in this paper.  However, if we average
the magnetic field strength and mass density over ever larger spherical
volumes centered on $r=0$, we do recover such a relationship: 
$\langle B \rangle
\propto \langle \rho\rangle^\kappa$, where angular
brackets denote the result of such a
spatial average and $\kappa = \Gamma/2$.

We may think of
the result $\langle B\rangle \propto \langle \rho\rangle^{\Gamma/2}$
as arising physically from a combination of two tendencies.
(a) Slow cloud contraction in the absence of magnetic fields
and rotation tends to keep roughly one Jeans mass inside every radius $r$,
which yields $\langle \rho\rangle \propto \langle c_s^2\rangle/G r^2$,
or $\langle \rho \rangle
\propto r^{-2/(2-\Gamma)}$ if $\langle c_s^2\rangle \propto \langle \rho
\rangle^{\Gamma -1}$.
(b) Slow cloud contraction in the absence of gas
pressure tends to keep roughly
one magnetic critical mass inside every radius $r$,
which yields $\langle B\rangle /r\langle \rho \rangle
\propto \lambda_r$ = constant,
or $\langle B \rangle \propto r\langle\rho\rangle \propto
r^{-\Gamma/(2-\Gamma)} \propto  \langle \rho\rangle^{\Gamma/2}$
if gas pressure (thermal
or turbulent) plays a comparable role to magnetic fields
in cloud support.  Notice that our reasoning does not rely
on arguments of cloud geometry, e.g., whether
cloud cores flatten dramatically or not as they contract; nor does it
depend sensitively on the precise reason for core
contraction, e.g., because of ambipolar diffusion or
turbulent decay.

Crutcher (1998) claims that the observational data are
consistent with $\kappa = 0.47 \pm 0.05$.  If we take Crutcher's
conclusion at face value, we would interpret the observations as
referring mostly to regions where the EOS is close to being isothermal
$\Gamma \approx 1$, which is the approximation adopted by many
theoretical studies that ignore the role of cloud turbulence.
The result is not unexpected for low-mass cloud cores,
but we would not naively have expected this relationship
for high-mass cores and cloud envelopes, where the importance
of turbulent motions is much greater.
Unfortunately, the observational data refer to different clouds rather
than to different (spatially averaged) regions of the same cloud, so
there is some ambiguity how to make the proper connection to different
theoretical predictions.  There may also be other
mechanisms at work, e.g., perhaps a tendency for
observations to select for regions of nearly constant
Alfv\'en speed, $\langle v_A \rangle \propto \langle B\rangle/\langle \rho
\rangle^{1/2} \approx$ constant
(Bertoldi \& McKee 1992).  Thus, we would warn the
reader against drawing premature conclusions about the effective EOS
for molecular clouds, or the related degree to which observations can
at present distinguish whether molecular clouds are magnetically
supercritical or subcritical.

If molecular clouds are magnetically supercritical, with $\lambda_r$
greater than 1 by order unity (say, $\lambda_r = 2$), then an
appreciable fraction (say, 1/2) of their support against self-gravity
has to come from turbulent or thermal pressure (Elmegreen 1978, McKee
et al. 1993, Crutcher 1998).  Modeled as scale-free equilibria, such
clouds with $\mu^2$ of order unity are not highly flattened (see Tables
1, 2 and Figs. 2, 3).  Suppose we try gravitationally to extract a
subunit from an unflattened massive molecular cloud, where the
cloud as a whole is only somewhat supercritical, $\lambda_r \sim 2$.
If the subunit's linear size is smaller than
the vertical dimension of the cloud by more than a factor of 2,
which will be the case if we consider subunits of stellar mass
scales, then this subunit will not itself
be magnetically supercritical.  Magnetically
subcritical pieces of clouds cannot contract indefinitely
without flux loss, so star formation in {\it unflattened}
clouds, if they are not highly supercritical, needs
to invoke some degree of ambipolar diffusion
in order to produce small dense cores that can gravitationally
separate from their surroundings.

On the other hand, if molecular clouds are
magnetically critical or subcritical, with $\lambda_r \le 1$, then
almost all scale-free equilibria are highly flattened, with $\mu^2$
appreciably less than unity.  On a small scale,
any subunit of this cloud, even subunits with vertical
dimension comparable to the cloud as a whole, would
also be magnetically critical or subcritical.
For such a subunit to contract indefinitely, we would again
need to invoke ambipolar diffusion to make a cloud core
magnetically supercritical.  Thus, although
the decay of turbulence can accelerate the formation
of cloud cores, the ultimate formation of stars from
such cores may still need to rely on {\it some} magnetic flux
loss (but perhaps not more than a factor of $\sim 2$)
to trigger the evolution of the cores toward gravomagneto catastrophe
and a pivotal state with a formally infinite central concentration.

On the large scale, if GMCs
are modeled as flattened isopedic sheets, \cite{SL97}
proved that magnetic pressure and tension are proportional to the gas
pressure and force of self-gravity.  Their theorems hold independently
of the detailed forms of the EOS or the surface density distribution.
If GMCs are truly highly flattened -- with typical
dimensions, say, of 50 pc $\times$ 50 pc $\times$ a few pc or even less
-- then many aspects of their magnetohydrodynamic stability
and evolution become
amenable to a simplified analysis through the judicious application and
extension of the theorems proved by \cite{SL97} (e.g., see Allen \& Shu
1998b, \cite{shuetal99}).  This exciting possibility
deserves further exploration.

\bigskip
\acknowledgments
D.G. acknowledges support by CNR grant 97.00018.CT02, ASI grant
ARS-96-66 and ARS-98-116, and hospitality from UNAM, M\'exico.  S.L.
acknowledges support by J. S. Guggenheim Memorial Foundation, grant
DGAPA-UNAM and CONACyT, and hospitality from
Osservatorio di Arcetri.  F.C.A. is supported by NASA grant
No.~NAG~5-2869 and by funds from the Physics Department at the
University of Michigan. The work of F.H.S. is supported in part by an
NSF grant and in part by a NASA theory grant awarded to the Center for
Star Formation Studies, a consortium of the University of California at
Berkeley, the University of California at Santa Cruz, and the NASA Ames
Research Center.

\clearpage

\begin{deluxetable}{llll}
\footnotesize
\tablecaption{\sc Parameters of Equilibria with $H_0=0.5$}
\tablewidth{0pt}
\tablehead{
\colhead{$\Gamma$} & \colhead{$\lambda_r$} & \colhead{$D$} & \colhead{$\mu^2$}\\
\colhead{} & \colhead{} & \colhead{} & \colhead{} 
}

\startdata
1     &   1.94  & 1.50 & 0.668  \nl
0.9   &   1.80  & 1.68 & 0.632  \nl
0.8   &   1.63  & 1.76 & 0.543  \nl
0.7   &   1.45  & 1.80 & 0.434  \nl
0.6   &   1.27  & 1.84 & 0.321  \nl
0.5   &   1.09  & 1.91 & 0.218  \nl
0.4   &   0.927 & 2.14 & 0.122  \nl
0.3   &   0.947 & 5.06 & 0.0165 \nl
0.2   &   0.992 & 13.8 & 0.00181 \nl
\enddata
\end{deluxetable}

\clearpage
 
\begin{deluxetable}{llll}
\footnotesize
\tablecaption{\sc Parameters of Equilibria with $\Gamma=1/2$}
\tablewidth{0pt}
\tablehead{
\colhead{$H_0$ } & \colhead{$\lambda_r$} & \colhead{$D$} & \colhead{$\mu^2$}\\
\colhead{} & \colhead{} & \colhead{} & \colhead{} 
}
 
\startdata
0     & $\infty$ & 1      & $\infty$ \nl
0.05  & 4.35     & 1.18   & 6.93     \nl
0.1   & 2.83     & 1.21   & 2.81     \nl
0.2   & 1.85     & 1.31   & 1.08     \nl
0.3   & 1.45     & 1.45   & 0.577    \nl
0.4   & 1.23     & 1.64   & 0.348    \nl
0.5   & 1.09     & 1.91   & 0.218    \nl
0.6   & 1.01     & 2.34   & 0.134    \nl
0.7   & 0.980    & 3.00   & 0.0786   \nl
0.8   & 0.977    & 3.91   & 0.0448   \nl
0.9   & 0.982    & 4.95   & 0.0265   \nl
1.0   & 0.987    & 6.06   & 0.0165   \nl
1.1   & 0.991    & 7.22   & 0.0108   \nl
1.2   & 0.993    & 8.44   & 0.00733  \nl
1.3   & 0.995    & 9.71   & 0.00514  \nl
1.4   & 0.996    & 11.04  & 0.00369  \nl
1.5   & 0.997    & 12.42  & 0.00272  \nl
\enddata
\end{deluxetable}
 
\clearpage

\begin{deluxetable}{llll}
\footnotesize
\tablecaption{\sc Parameters for the Thin Disk Limit}
\tablewidth{0pt}
\tablehead{
\colhead{$\Gamma$ } & \colhead{$\alpha$} & \colhead{$f(\pi/2)$} & 
\colhead{$R_0$}\\
\colhead{} & \colhead{} & \colhead{} & \colhead{}
}
 
\startdata
1     & $45^\circ$ & 1        & 2        \nl
0.9   & $52^\circ$ & 1.19     & 2.65     \nl
0.8   & $58^\circ$ & 1.52     & 3.61     \nl
0.7   & $64^\circ$ & 2.05     & 5.10     \nl
0.6   & $69^\circ$ & 2.93     & 7.55     \nl
0.5   & $73^\circ$ & 4.50     & 11.9     \nl
0.4   & $77^\circ$ & 7.59     & 20.5     \nl
0.3   & $81^\circ$ & 14.7     & 40.4     \nl
0.2   & $84^\circ$ & 36.8     & 102      \nl
0.1   & $87^\circ$ & 166      & 467      \nl
0     & $90^\circ$ & $\infty$ & $\infty$ \nl
\enddata
\end{deluxetable}

\clearpage

\label{fig1}
\figcaption[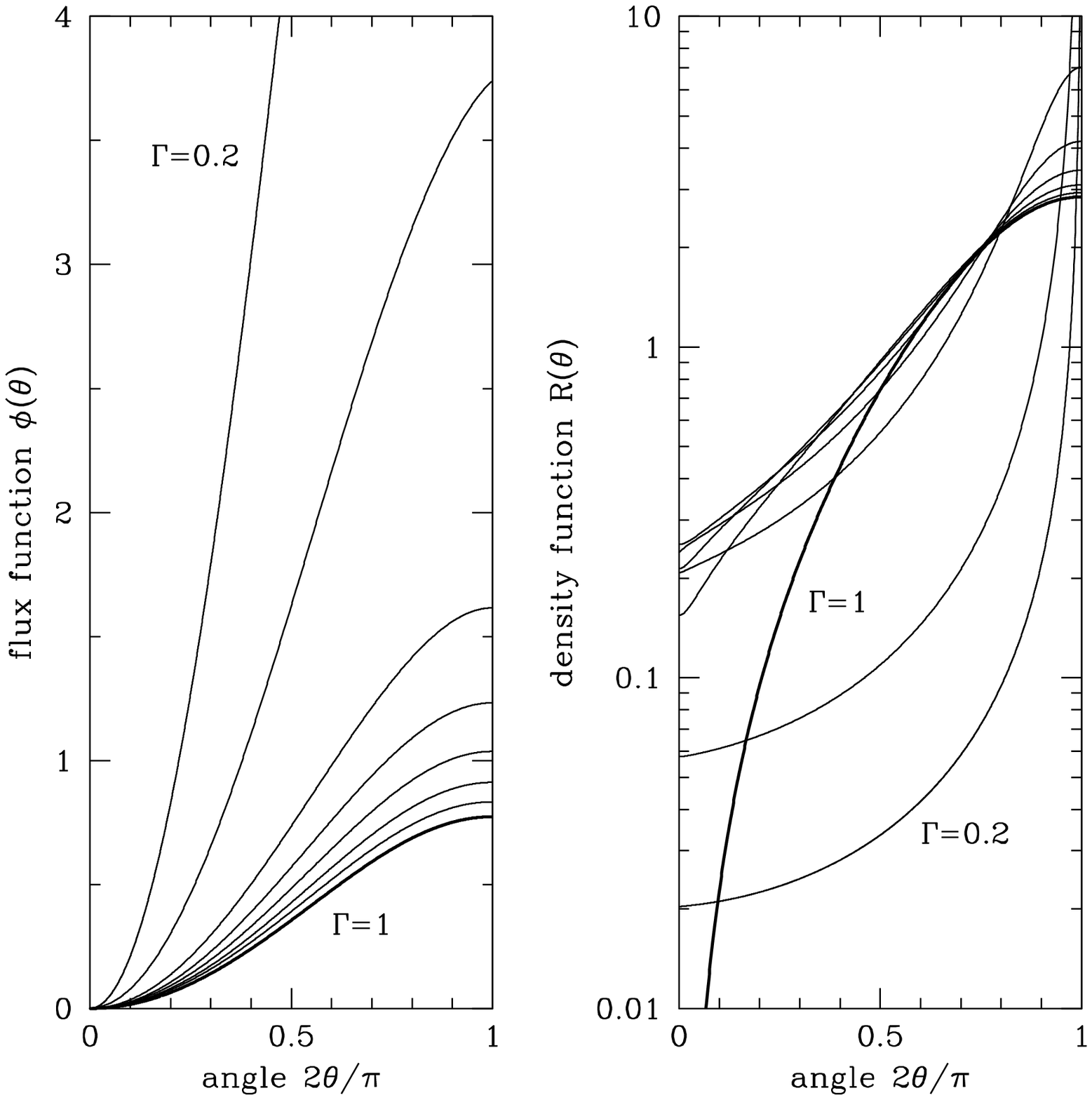]
{The flux and density functions $\phi(\theta)$ and $R(\theta)$ computed
for $H_0 = 0.5$ and $\Gamma=$ 1, 0.8, 0.7, 0.6, 0.5, 0.4, 0.3, and 0.2.
The isothermal case is indicated by {\em thick} lines.}

\label{fig2}
\figcaption[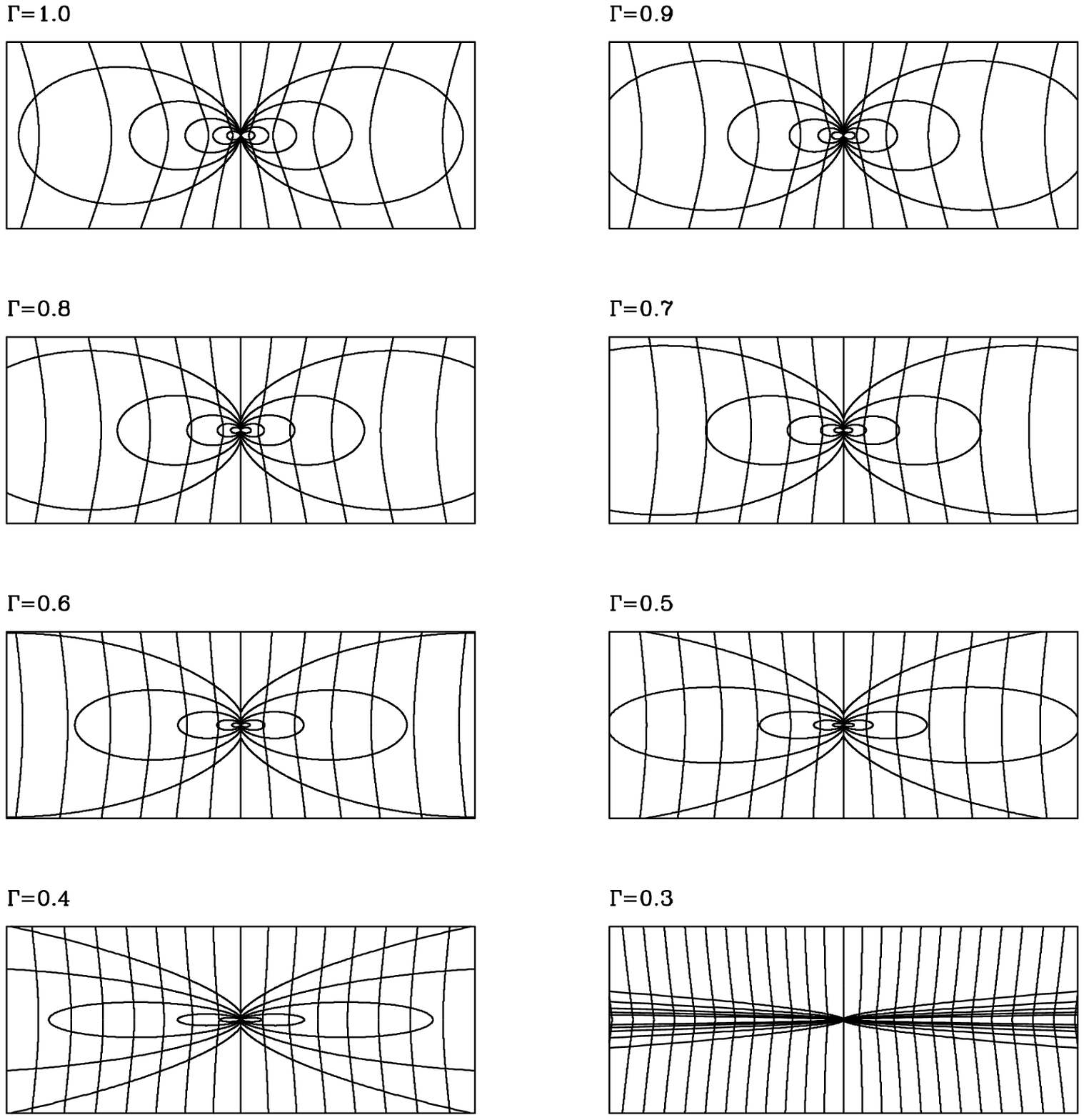]
{Density contours and magnetic field lines for $H_0=0.5$ and $\Gamma=$
1, 0.8, 0.7, 0.6, 0.5, 0.4, 0.3, and 0.2. The isodensity levels
correspond to $R(\theta) = 2^k$, the isoflux contours correspond to
$\phi(\theta) = (0.2 k)^2$, where $k=$ 0, 1, 2, $\ldots$.}

\label{fig3}
\figcaption[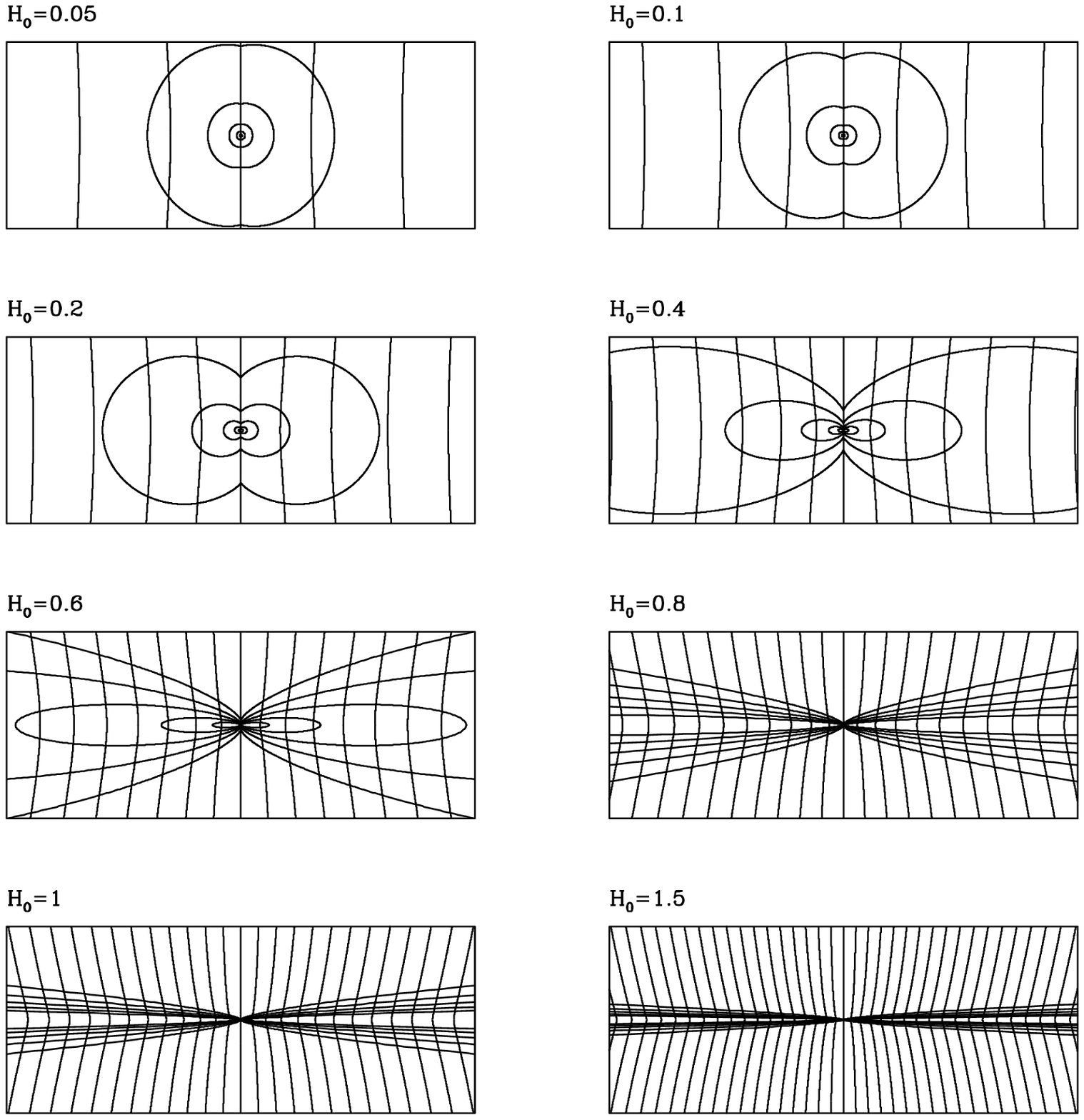]
{Density contours and magnetic field lines corresponding
to $\Gamma=1/2$ and $H_0=$ 0.05, 0.1, 0.2, 0.4, 0.6, 0.8, 1.0, and 1.5.
The isodensity and isoflux contours are the same as in Fig.~2.}

\label{fig4}
\figcaption[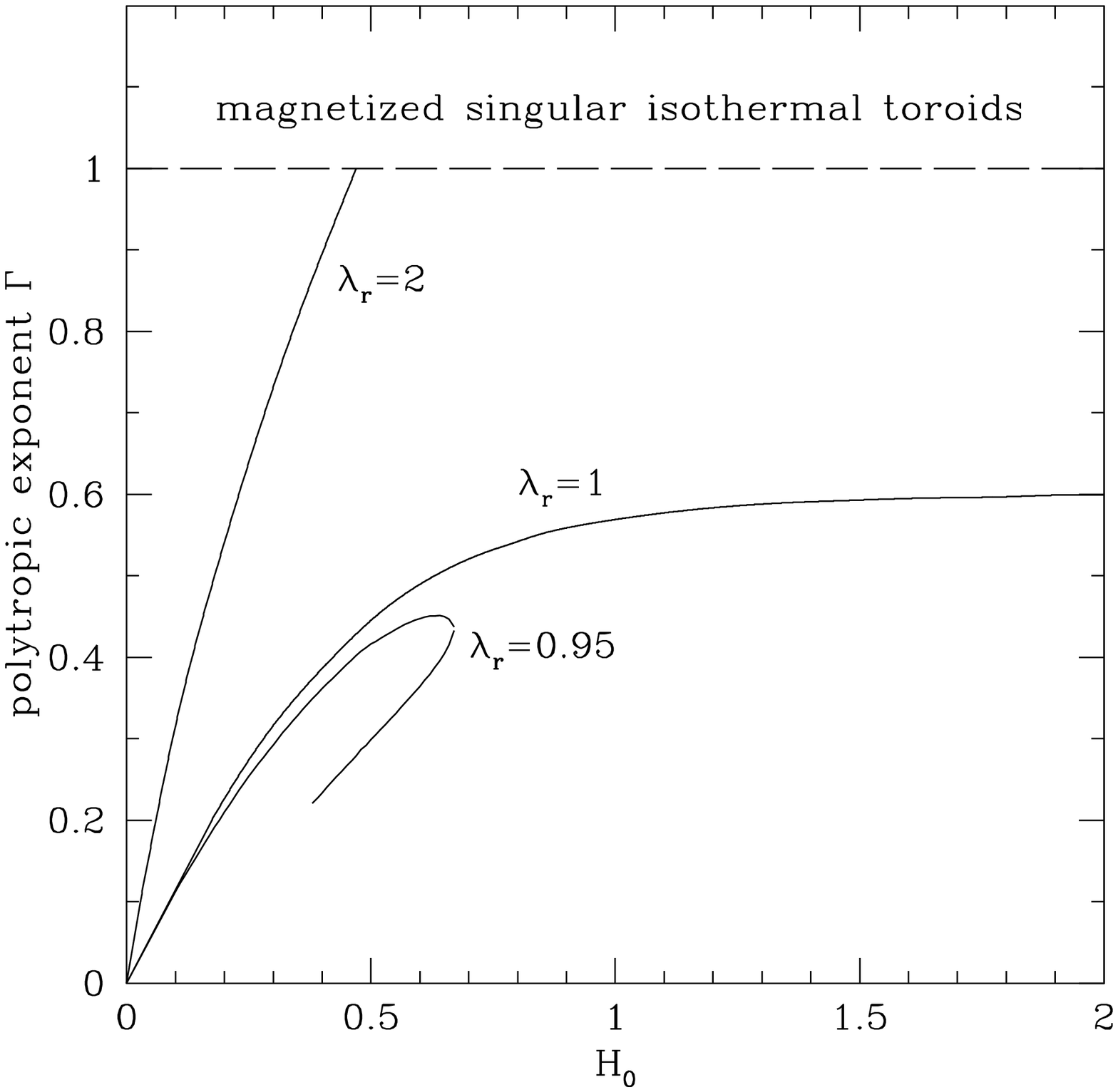]
{The locus of the set of equilibria with $\lambda_r=$ 0.95, 1, and 2 in the
($H_0$, $\Gamma$) plane.}

%\label{fig5}
%\figcaption[fig5.eps]
%{The asymptotic ($H_0 \gg 1$)  flux and density functions $f(\theta)$,
%$s(\theta)$,  for $\Gamma = 1, 0.9, 0.8, 0.7, 0.6, 0.5, 0.4, 0.3, 0.2, 0.1$.}

\label{fig5}
\figcaption[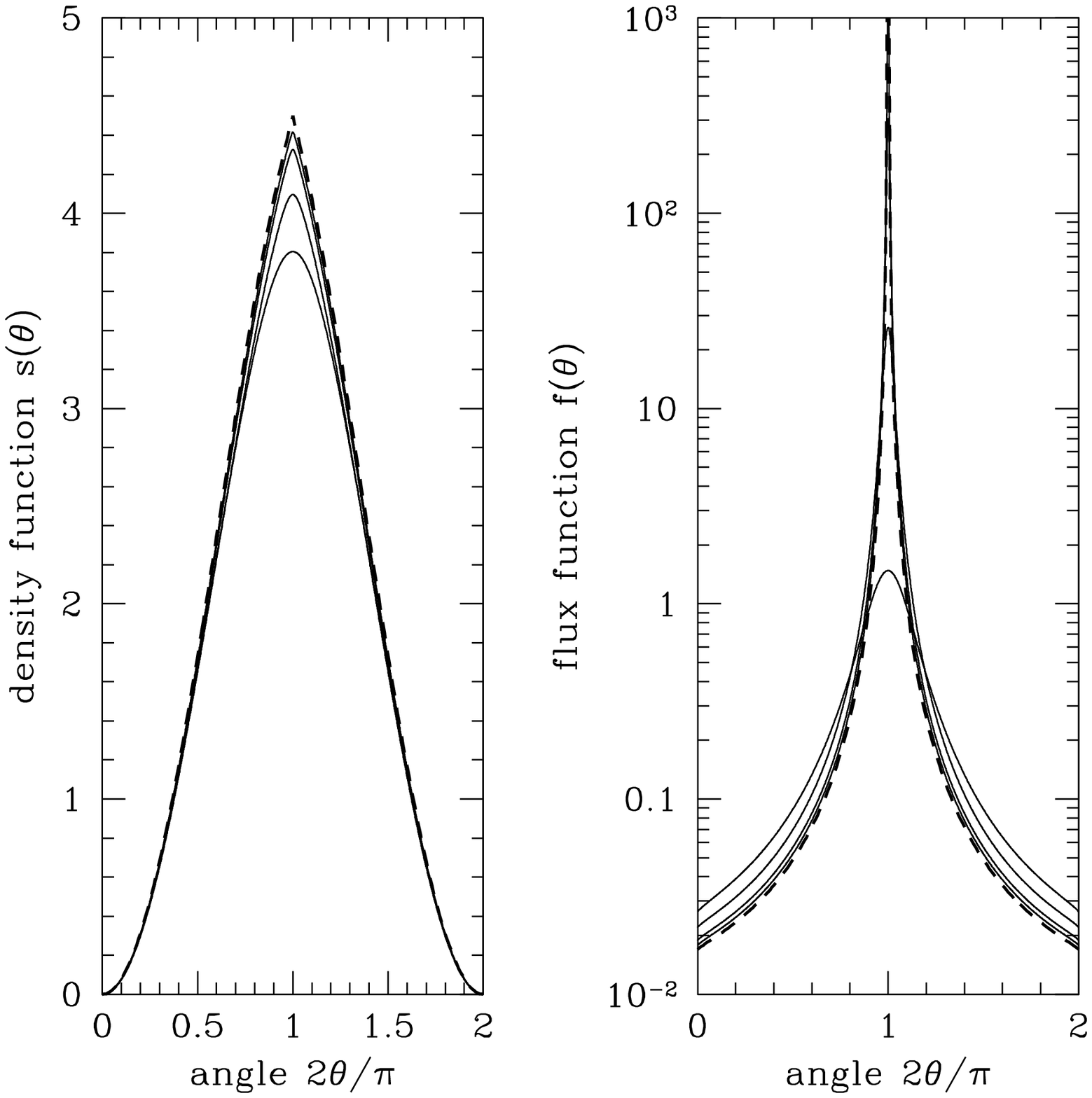]
{The functions $f(\theta)$ and $s(\theta)$ for $\Gamma=1/2$ and $H_0=$
0.4, 0.6, 0.8, 1.0, 1.2, and 1.5, compared with the asymptotic
expressions obtained for $H_0\gg 1$, shown by the {\em thick dashed}
lines. For clarity, the functions are shown over the range $0\le \theta
\le\pi$.}

%\label{fig7}
%\figcaption[fig7.eps]
%{The asymptotic ($H_0 \ll 1$) flux and density functions $g(\theta)$,
%$p(\theta$), for $\Gamma=1,0.9,0.8,0.7,0.6,0.5,0.4,0.2,0.2,0.1$. 
%The isothermal case is indicated by {\em thick} lines.}

\label{fig6}
\figcaption[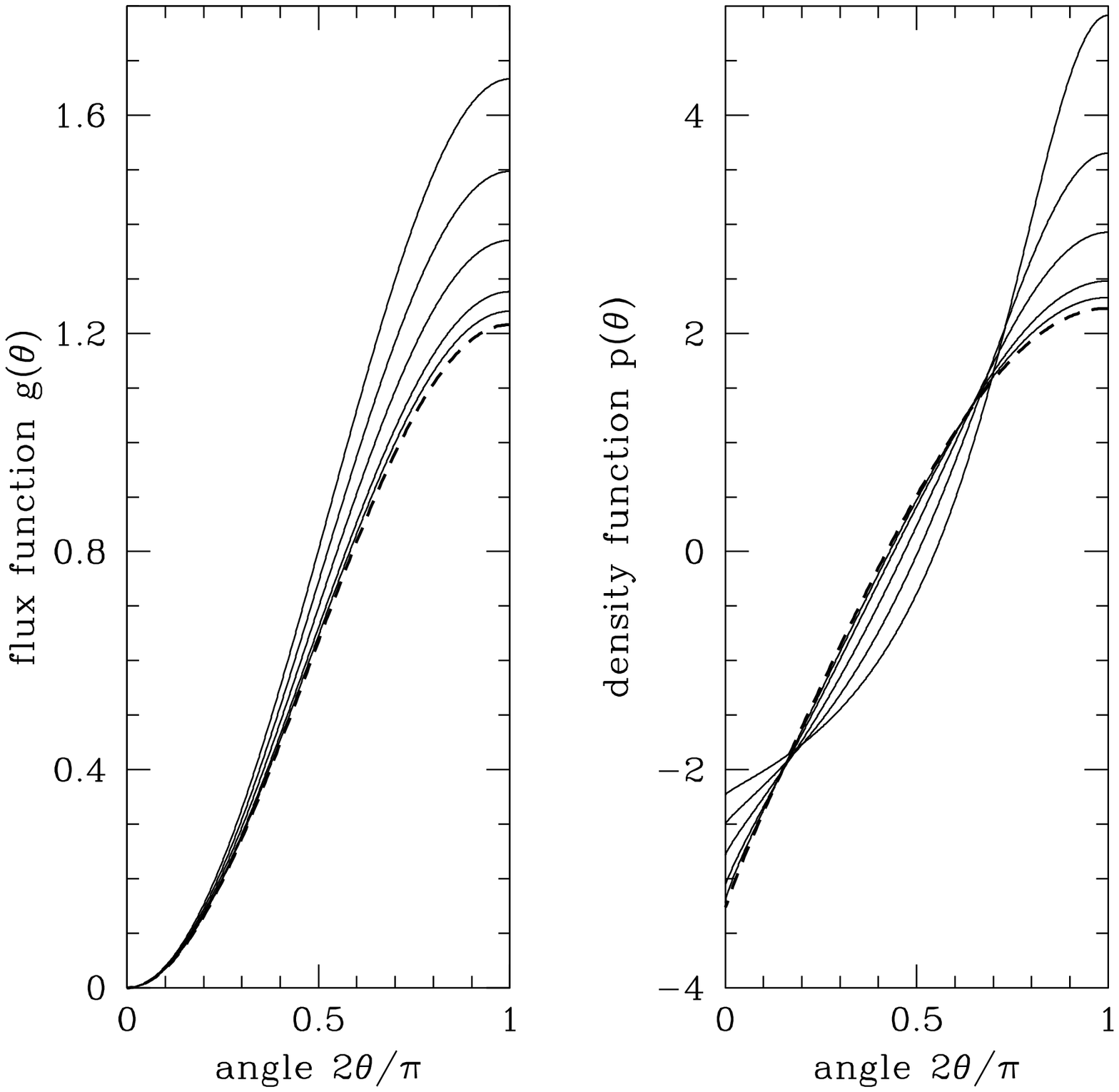]
{The functions $g(\theta)$ and $p(\theta)$ for $\Gamma=1/2$ and
$H_0=$ 0.4, 0.3, 0.2, 0.1, and 0.05, compared with the asymptotic expressions
obtained for $H_0\ll 1$, shown by the {\em thick dashed} lines.}

\label{fig7}
\figcaption[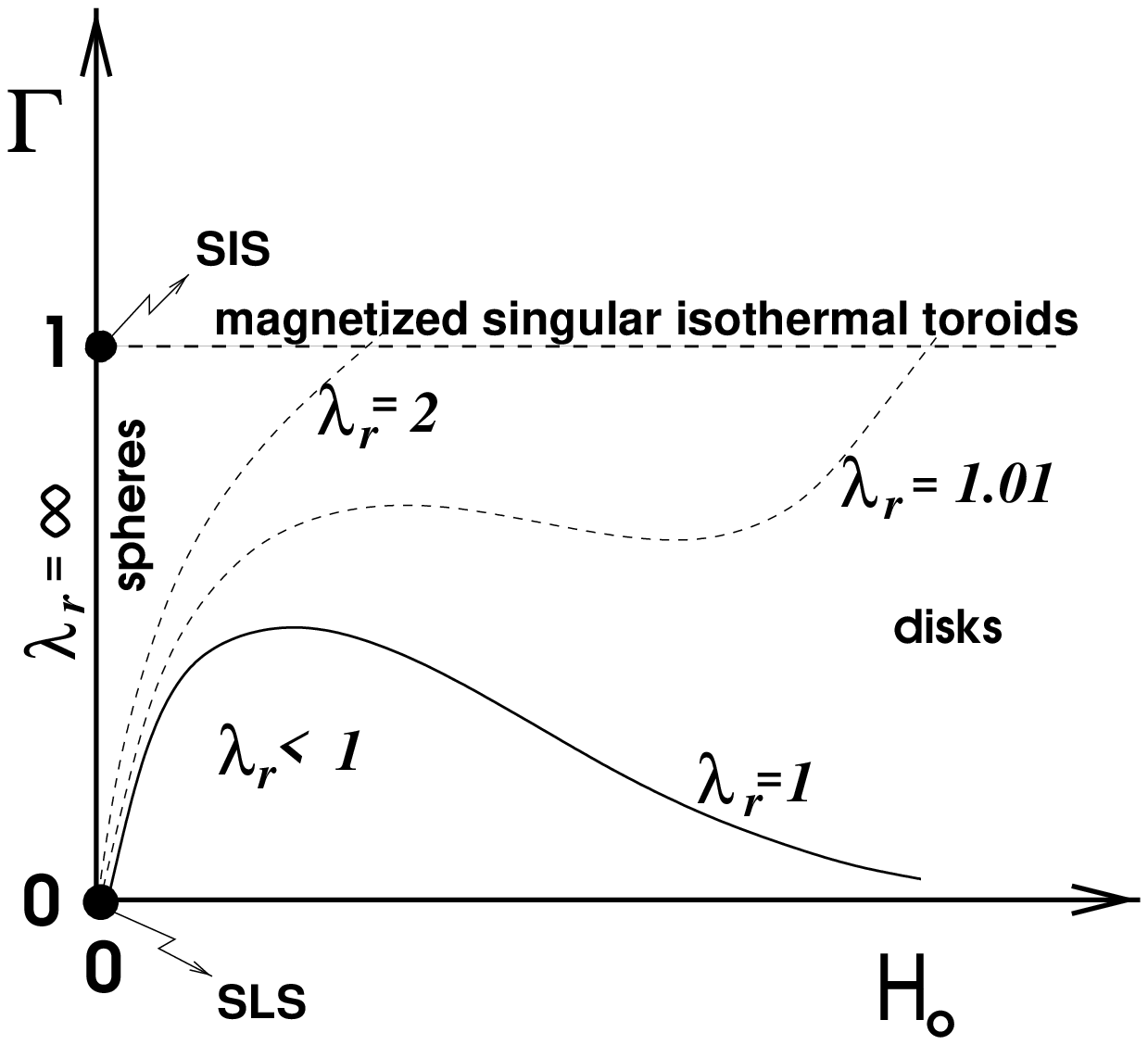]
{Schematic picture of the $(H_0, \Gamma)$ plane indicating the topology
of the solutions for scale free magnetized isopedic singular
self-gravitating clouds.}

\end{document}